\documentclass[global,twocolumn,nospthms]{svjour}

\usepackage{amsmath}
\usepackage{upgreek}
\usepackage{graphicx}
\usepackage[numbers]{natbib}

\journalname{}

\begin{document}
\sloppy

\title{High extinction amplitude modulation in ultrashort pulse shaping}
\author{Yen-Wei Lin \and Brian C. Odom}
\institute{Department of Physics and Astronomy, Northwestern University\\ 2145 Sheridan Road, Evanston, IL 60208}

\date{\today}

\maketitle

\begin{abstract}
We explored the issues related to the resolution and the modulation extinction when filtering the spectrum of a UV femtosecond laser with a standard ultrashort pulse shaper.  We have learned that a higher pulse shaping resolution often requires a larger working beam size or  a higher density grating for greater dispersion. However, these approaches also introduce more optical errors and degrade the extinction. In this work, we examined specifics of each component to determine the best configuration of our spectral filtering setup. As a proof-of-concept demonstration, we utilized elements available as standard products and achieved 100 GHz filtering resolution with high extinction at the UV-A wavelength, which is superb in this wavelength range. The high extinction spectral filtering is especially important while modifying a broadband laser for the optical control of molecule's internal state.
\end{abstract}

\section{Introduction}
The coherent broadband nature of ultrashort pulsed lasers greatly increases the bandwidth of interaction between the laser and a subject. Ultrashort pulse shaping technology further advances the degree of such interaction in many applications. For instance, in optical communication, pulse shaping technology is used to encode information and perform computation\citep{sardesai1998femtosecond}. It is also used to correct and optimize the phases of the spectrum for generating high-quality ultrashort pulsed lasers \citep{Pastirk:2006aa}. In controlling chemistry, an ultrashort pulse laser with a programmed waveform provides delicate control of molecules \citep{goswami2003optical,Sauer:2007aa}. 

\begin{figure}
\includegraphics[width=3in]{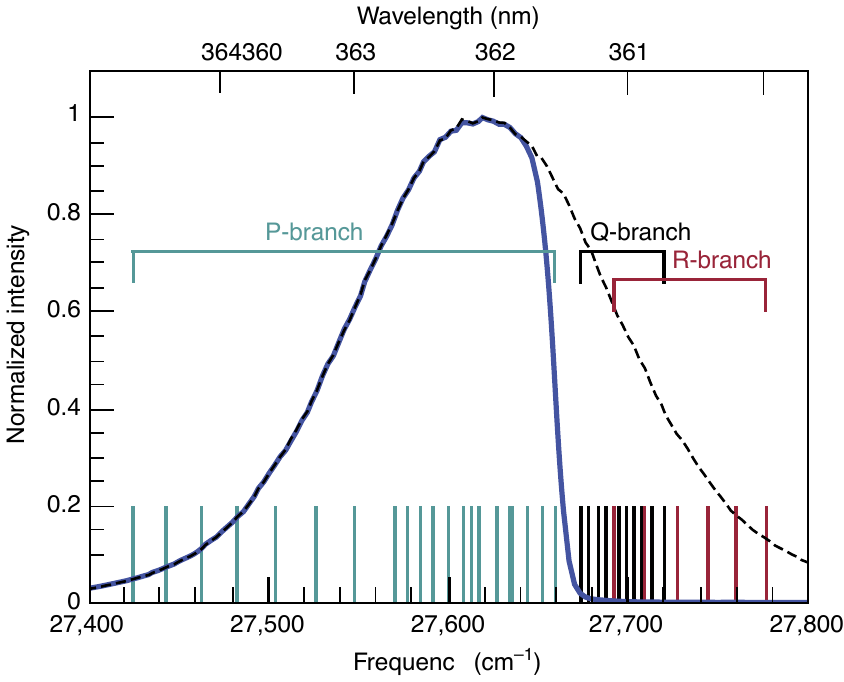}
\caption{Rotational cooling with spectral-filtered broadband source. In this example, relevant transitions from $\text{A}{}^2\Uppi_{1/2},\,v'=0 \leftarrow \text{X}{}^2\Upsigma_{1/2},\,v''=0$ band in the AlH$^+$ are shown as vertical bars at the bottom. P-branch is the rotational cooling transitions, while Q- and R-branch are the opposite. An unfiltered UV femtosecond laser (spectrum shown as dashed line) drives all transitions thus provides no cooling effect. A filtered source (spectrum shown as solid line) drives only cooling transitions and hence slows down the molecule's rotation.}\label{fig:demo}
\end{figure}

In our laboratory, we have been applying the pulse shaping technique to our molecular rotational state cooling experiment\citep{Lien:2014aa, Lien:2011aa}. The principle of rotational cooling is the optical pumping of transitions associated with losing rotational angular momentum (P-branch) while avoiding those diffusing (Q-branch) or heating (R-branch) the rotational population. This can be done by shaping a femtosecond laser's spectrum with a filter of which the P-branch is in the passband and the Q- and R-branch is in the stopband, as illustrated in Fig.~\ref{fig:demo}. The cutoff width of the filter determines how well one can avoid driving an unwanted transition, and ultimately determines the molecular cooling efficiency.

Ultrashort pulse shaping is the optical version of a waveform synthesizer. A typical implementation is similar to an optical spectrometer. The setup first Fourier transforms the input laser to obtain its optical spectrum with a dispersive element. Then, the angularly dispersed light is focused by a lens (or a concave mirror) to map the dispersion onto the focusing plane, where the spectrum is altered by a light modulator. Finally, the inverse Fourier transform is performed with another set of lens and dispersive element to form the output laser pulse with the desired spectrum or time-domain waveform.

High-resolution pulse shaping techniques have been the focus of ultrafast laser research for over two decades. Usually, the pulse shaping frequency resolution $\delta\nu$ is compared to the bandwidth $\Delta\nu$ of the unshaped laser as the figure of merit; the ratio $\eta=\Delta\nu/\delta\nu$ measures the number of pulse shaping control parameters. However, in our research where the laser is used to drive transitions, it is necessary to compare $\delta\nu$ to the molecular structure. Therefore, we emphasize the absolute frequency resolution rather than the fractional. One of the earliest pulse shaping implementations achieved $\approx$ 0.2 nm (51 GHz) resolution at 1064 nm\citep{Thurston:1986aa}. For the Ti:Sapphire laser wavelength (800 nm), modulation at around 0.1 nm (48 GHz) has been reported in \citep{Monmayrant:2004aa, Stobrawa:2001aa}. Recently, with the introduction of virtually imaged phase arrays (VIPAs), pulse shaping reached a few GHz resolution at similar wavelength\citep{Lee:2006aa, Supradeepa:08}. In the UV regime, there was, however, slower progress due to the lack of suitable light modulators. Yet, sub-nm resolution ($>$200 GHz) has been demonstrated at some UV wavelengths\citep{Weber:2010aa}. In all these experiments, pulse shaping resolution does not necessarily include information about modulation extinction contrast. The cutoff feature of the amplitude modulation was seldom mentioned. For cooling molecules with a pulse-shaped laser to work for heavier molecules such as oxides or fluorides, resolution better than 100 GHz with high extinction ratio is required and is the main goal of this work.

In this paper, we explore spectral filtering of a near UV femtosecond laser by providing an introductory element-wise investigation of the setup. We first briefly review the theory of ultrashort pulse shaping in the next section. Then, we investigate each component used in our setup in Sec.~\ref{sec:other_elements}. The evaluation and calibration of our spectral filtering are in Sec.~\ref{sec:evaluation}. In passing, this work is towards our molecular cooling experiment with SiO$^+$, for which some parameters are picked. Specifically, the relevant transitions are at 385 nm, and the rotational heating and the cooling transitions are simply separated by an 84 GHz gap.

\section{The spectral filtering setup}\label{sec_intro_filtering}
\begin{figure}
\includegraphics[width=3in]{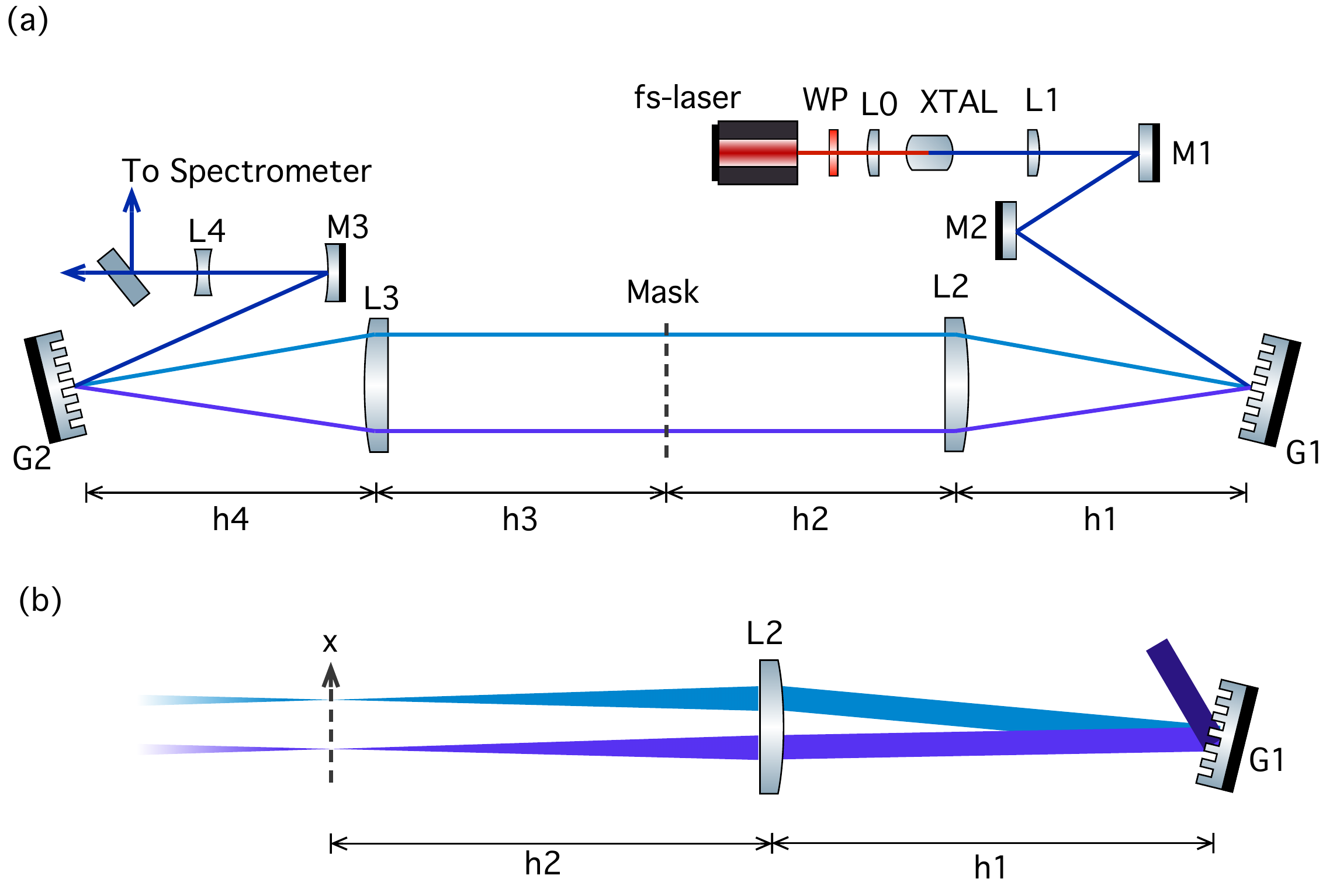}
\caption{Schematic of our spectral filtering setup. fs-laser: mode-locked femtosecond laser; WP: half-wave plate; XTAL: BBO SHG crystal; L1-4: lens; M1-3: mirrors (M3 is concave). G1 and G2 are reflective diffraction gratings; C: fiber coupler. See text for detailed discussion.}\label{fig_setup1}
\end{figure}

Our broadband spectral filtering setup is shown in Fig.~\ref{fig_setup1}(a). The unfiltered laser source is coupled into the setup by folding mirrors M1 and M2. Grating G1 and lens L2 Fourier transform the light and produce the laser spectrum. Specifically, the grating separates each spectral component into different angles. When focused by the lens, each component becomes a spot at different locations on the focusing plane, which makes the (optical) frequency plane. See Fig.~\ref{fig_setup1}(b). Filtering is achieved by masking the spectrum at the plane. The filtered spectrum is recombined into a single beam by L3 and G2, which performs the inverse Fourier transformation. The output beam is further reshaped with a telescope (M4 and L4) to a proper size for future experiments.

The ultrashort pulse shaping technique has been greatly reviewed in several articles including \citep{Monmayrant:2010aa,Weiner:2011aa}. Here we highlight the key results. Consider a Gaussian input beam with $1/e^2$ beam radius $w_i$. A spectral component of wavelength $\lambda$ makes a focused spot with beam radius
\begin{equation}\label{eqn_beamsize}
w_o = \frac{\cos\theta_i}{\cos\theta_d}\frac{f\lambda}{\pi w_i}
\end{equation}
at the frequency plane. In the above equation, $f$ is the focal length of the lens; $\theta_i$ and $\theta_d$ are the incident and diffraction angle at the grating, respectively. The dispersion at the frequency plane is
\begin{equation}\label{eqn_dispersion}
\frac{dx}{d\nu}=\frac{\lambda^2f}{cd\cos\theta_d}
\end{equation}
where $d$ is the pitch of the grating groves. The pulse shaping resolution is then the frequency extent across the spot:
\begin{equation}\label{eqn_resolution}
\delta\nu = w_o\times\frac{d\nu}{dx} = \frac{cd\cos\theta_i}{\pi\lambda w_i}.
\end{equation}

As a numerical example, we consider an input beam with $\lambda=385\;\text{nm}$ and $w_i=5\;\text{mm}$. We assume $\theta_i = \theta_d - 10^\circ = 49.1^\circ$. With $f = 500\;\text{mm}$ lens and 2400 groves/mm grating, the focus size is $w_o= 15.6 \;\mu\text{m}$ and the frequency resolution is $\delta\nu= 13.5\;\text{GHz}$. In this example, the theoretical spectral filtering resolution will be sufficient for the SiO$^+$ cooling experiment.
The resolution is entirely set by the grating density and the input beam size, independent to the imaging lens. However, the focus size and the dispersion are proportional to the focal length $f$. If a lens with shorter $f$ is used, the smaller spot size becomes difficult to mask in terms of mechanical resolution. On the other hand, using a longer $f$ reduces the mechanical challenge but increases the setup's footprint and requires larger optics, which makes optical aberration become an issue. Our choice of $f=500\;\text{mm}$ represents a compromise between the above two concerns, along with other matters discussed later.

The theoretical treatment of an ultrashort pulse system by the Wigner function is a well-established method \citep{Paye:1995aa,Wefers:1996aa}. On the other hand, the response of the optics can be described by the optical transfer function (OTF). The combination of these two thus provides the comprehensive theoretical study of a real world pulse shaper. However, most of the previous works adopted the paraxial approximation such that aberrations were not considered at all. Here, we only point out but skip the theoretical investigation as it is outside the scope of our experimental work.

\section{Setup}\label{sec:other_elements}
In this section, we introduce and comment on components in our spectral filtering setup as well as the alignment procedure.

\subsection{Light source}\label{sec:light_source}
The input 385 nm broadband source is the second harmonic generation (SHG) of a mode-locked femtosecond laser (Spectra Physics Mai-Tai). This is done by focusing the fundamental (770 nm) light with an f = 30 mm achromatic lens onto a thin (0.5 mm) BBO crystal\citep{Zhang:1998aa, Ghotbi:2004aa, Ghotbi:2004ab}. The UV beam from the crystal is collimated by a doublet lens with approximately 400 mm effective focal length. We found it is sufficient to compare the beam size at near and far fields to determine collimation. The collimated UV beam's radius is 4.8 mm beam horizontally and 5.7 mm vertically. The results were obtained by the knife edge beam size measurement. The data presented an acceptable Gaussian intensity distribution.

We noticed that the UV beam acquires dispersion from the SHG process, most likely due to the walk-off. We then set up the SHG such that the dispersion is aligned to the grating's dispersion direction, and accommodate such initial dispersion with the pulse shaper. Further discussion continues in Sec.~\ref{subsec:dispersion_compensation}.

\subsection{Grating}
We use holographic gratings with 2400 groves/mm in our spectral filtering setup. These gratings are set up in a close to Littrow configuration: the incident angle is $\theta_i=47.1^\circ$ and the diffraction angle is $\theta_d=40.9^\circ$ at $\lambda=385\;\text{nm}$.
While our choice of the grating is due to component availability, a grating with higher density is of course preferred. We would avoid ruled gratings as they are prone to periodic and irregular ruling errors that result in ghosting and stray light and hurt the filtering resolution. However, blazed ruled gratings can be more efficient; our holographic grating has about 1/2 diffraction efficiency and is the least efficient component in the setup.

\subsection{Focusing lens}\label{sec:lens}
A large aperture lens is used to accommodate the broadband light, which is a few centimeters wide at L2. In such case, all types of optical aberration can be easily picked up and makes the focus less tight. Spherical aberration is the dominant type. For tilted incident, astigmatism is somewhat less problematic in one-dimensional pulse shaping. Coma, on the other hand, produces a comet-like spot shape that indeed corrodes the resolution in an asymmetric fashion. Lastly, we safely ignore chromatic aberration since $\Delta \lambda\approx 2\;\text{nm}<\lambda =385\;\text{nm}$.
Although there is a quality spectrum of stocked optics available from visible to IR wavelength, off-the-shelf optics for our application in UV is quite limited. The option to be considered is the combining of spheric lenses into a doublet lens to reduce all types of aberration mentioned above. Constructing a doublet lens usually involves using two different glasses and optimizing the curvatures of each surface. While we have access to only some 2" fused silica lenses, our doublet lens is constructed by stacking a positive meniscus lens (Thorlabs LE4822) on a plano-convex lens (Thorlabs LA4337). The effective focal length of this doublet is around 500 mm.

\subsection{Mask}
\begin{figure}
\includegraphics[width=3in]{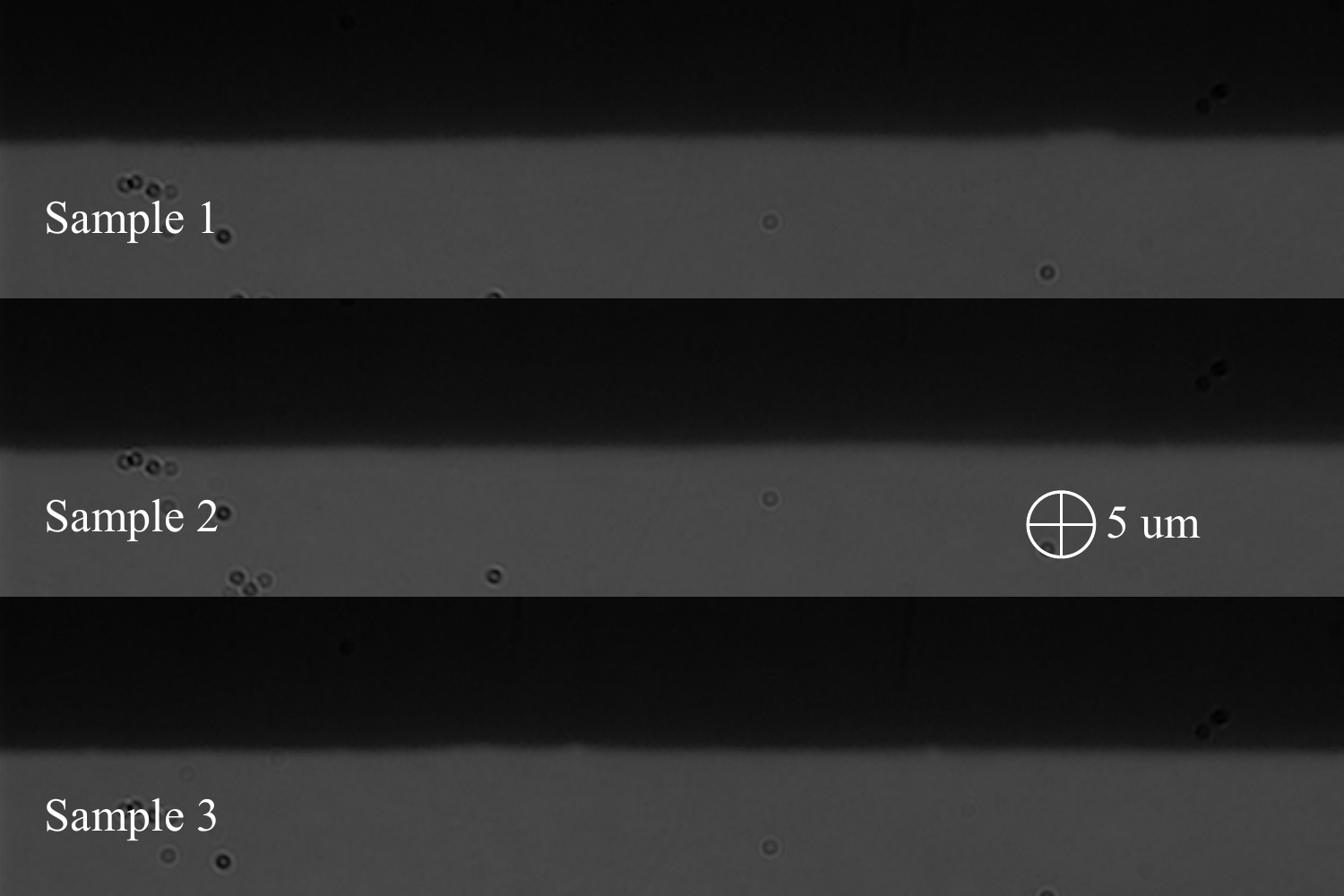}
\caption{Microscope images of three samples of razor blades used in this experiment. While we image the blades (gray area) in front of a dark background, the boundary between two areas is due to the blade edge. It is easy to see the fuzziness of the blade edge is around 1 $\upmu\text{m}$, smaller than the focal spot size ($\sim 10\;\upmu\text{m}$) in our setup.}\label{fig_blade_image}
\end{figure}

The SiO$^+$ rotational cooling requires a simple low-pass filtering mask, which is realized by a razor blade. Installing the mask is straightforward as we place the mask at the location where best resolution is obtained. That is, the longitudinal location of the blade is searched, while the transverse one is set to a specific cutoff frequency based on the molecule. The knife edge is set vertically but a little tilt is not too critical as it will soften the cutoff by $1/\cos\theta$, which is tiny for small tilt angle $\theta$ .
We inspected the razor blades under a microscope (see Fig.~\ref{fig_blade_image}) and found typical edge roughness is of order 1 $\upmu$m - smaller than the focal spot size. Hence the quality of the blade edge is not yet the limiting factor to our spectral filtering resolution. But we would avoid a used blade because of the likely worn knife edge.

\subsection{Dispersion compensation}\label{subsec:dispersion_compensation}
The output beam acquires dispersion because the second grating G2 does not properly recombine the filtered spectrum. The mathematical description follows. We start with the grating diffraction equation at G2:
\begin{equation}
d (\sin\theta_{i}+\sin\theta_{d})=\lambda.
\end{equation}
By expanding the above equation to the first order in $\delta \theta_{i}$, $\delta \theta_{d}$, and $\delta \lambda$, we obtain the first order dispersion:
\begin{equation}\label{eqn_g1}
\frac{\delta\theta_{d}}{\delta\lambda}=\frac{1}{d\cos\theta_{d}^0}-\frac{\cos\theta_{i}^0}{\cos\theta_{d}^0}\frac{\delta\theta_{i}}{\delta\lambda}
\end{equation}
with
\begin{equation}
d(\sin\theta_{i}^0+\sin\theta_{d}^0)=\lambda_0
\end{equation}
being the diffraction condition at the center wavelength $\lambda_0$. The term $\delta\theta_{i}/\delta\lambda$ on the right hand side of Eq.~\ref{eqn_g1} is used to characterize the propagation of filtered spectral components when they merge onto the grating. The output beam is dispersion-free when $\delta\theta_{d}/\delta\lambda =0$ and is fulfilled at a specific grating angle
\begin{equation}
\theta_{i}^0 = \cos^{-1}[\frac{1}{d}(\frac{\delta\theta_{i}}{\delta\lambda})^{-1}].
\end{equation}

\begin{figure}
\includegraphics[width=3in]{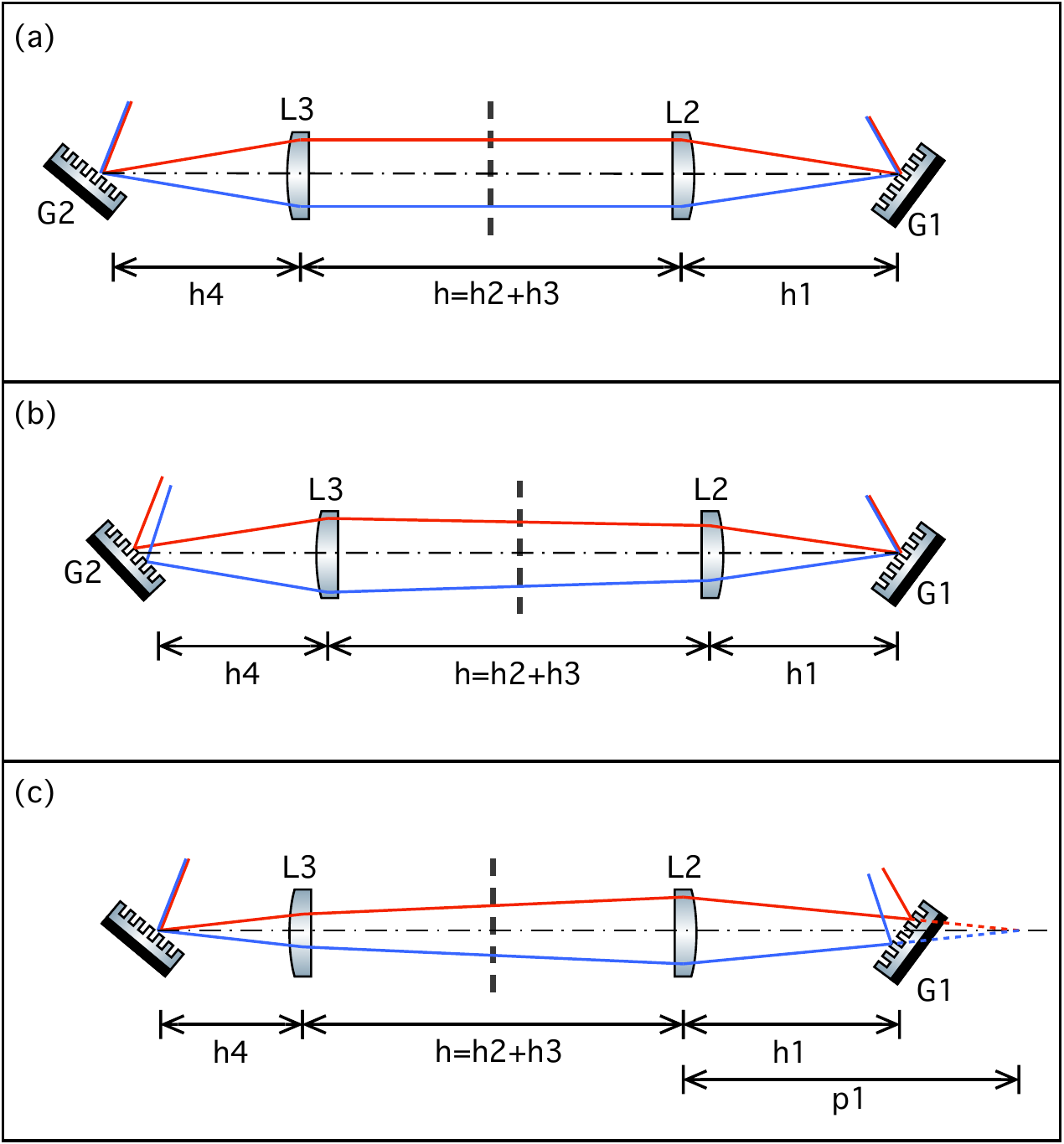}
\caption{Some possible scenarios regarding dispersion in the 4-f line. The red and the blue lines represent the propagation path of two different spectral components. (a) is an ideal case where the input beam is dispersion-free and G1=G2 and L2=L3. In (b), the alignment of some components is off, and hence the output beam derives dispersion. In general, as in (c), the source of dispersion can come from the input beam and errors from components and alignment. However, dispersion can be compensated by properly aligning the location and the diffraction angle of the grating G2.}\label{fig:dispersion}
\end{figure}

Qualitatively, Fig.~\ref{fig:dispersion} illustrates some possible scenarios. (a) is the flawless case where the input beam has no dispersion and the optics are matched pairs. In this case, a symmetric 4-f line is aligned and can be reduced to a 2-f configuration by retro-reflecting the filtered spectra at the frequency plane. In (b), the distance $h1$ is slightly off $f$ such that each spectral component does not propagate in parallel between L2 and L3. This angular error eventually results in the output beam's dispersion if G2 is not adjusted correspondingly. (c) shows the general situation where the source of error includes the input laser already carrying dispersion and optical elements not being identical. As explained earlier, the sum of all these errors is compensated by optimizing the location and the angle of the recombining grating G2. It is worth mentioning that the input laser's dispersion alone can be corrected beforehand. Here, we simply delay that task as we can correct for some other errors as well. Other than this aspect, the 4-f configuration is expected to have equal performance as a 2-f configuration.

\subsection{Alignment}\label{subsec:alignment}
We align elements in the 4-f line sequentially. After grating G1, the focusing lens L2 is set approximately one focal length away without further optimization. The location of L3 is adjusted such that the two lenses (L2 and L3) form a 1:1 Galilean telescope.

The recombining grating G2 should be placed where all spectral components merge; its angle is tuned to diffract all components into the same direction. To check the alignment, we use a mask with two tiny slits to select only components from both ends of the spectrum. We then optimize the grating until these spectral components form an overlapped output. For example, if one sees two spots, the location of the grating needs adjustment. Furthermore, when two spots travel at crossing paths, we know the diffraction angle is off. After G2 is optimized, the recombined output beam remains a single spot over a long propagation distance.

\begin{figure}
\includegraphics[width=3in]{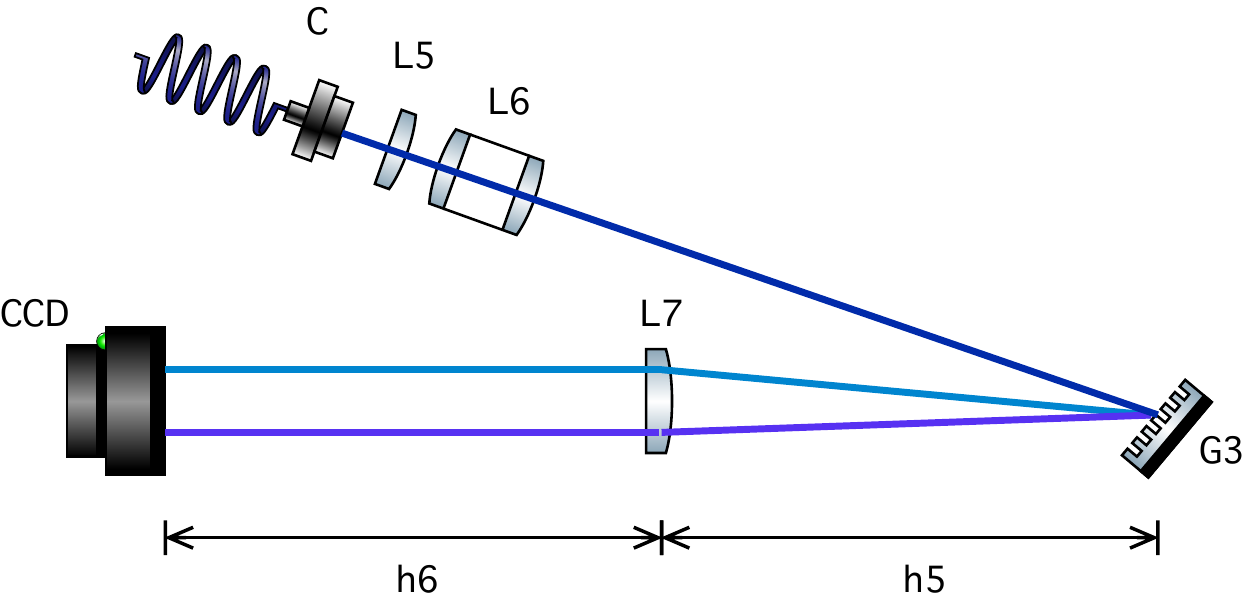}
\caption{Schematic of our spectrometer. C: fiber output coupler; L5-6: collimation lens; G3: diffraction grating; L7: focusing lens; CCD: linear camera.}\label{fig_setup2}
\end{figure}

\section{evaluation and calibration}\label{sec:evaluation}

\subsection{Spectrometer}
In order to diagnose our spectral filtering, we also built a high-resolution spectrometer as outlined in Fig.~\ref{fig_setup2}. The input light is coupled into the spectrometer by a single mode fiber and is collimated to 6.8 mm beam radius. We use 3600 grooves/mm ruled grating in our spectrometer. The focusing lens L5 is the same as the one in spectral filtering. A linear camera is placed on the frequency plane to record the spectrum; it has 3648 pixels and each is 8 $\upmu$m wide and 200 $\upmu$m tall. Based on the data presented later, the $1/e^2$-full-width frequency resolution of this spectrometer is 28 GHz.

Our homebuilt spectrometer was also the test bed for the focusing lens. The evaluation incorporated a narrowband CW laser to produce a single spectral component on the frequency plane. We took this approach to check the performance of our homebuilt doublet lens.

\subsection{Permissible input beam size}\label{sec_evaluate1}
\begin{figure}
\includegraphics[width=1.5in]{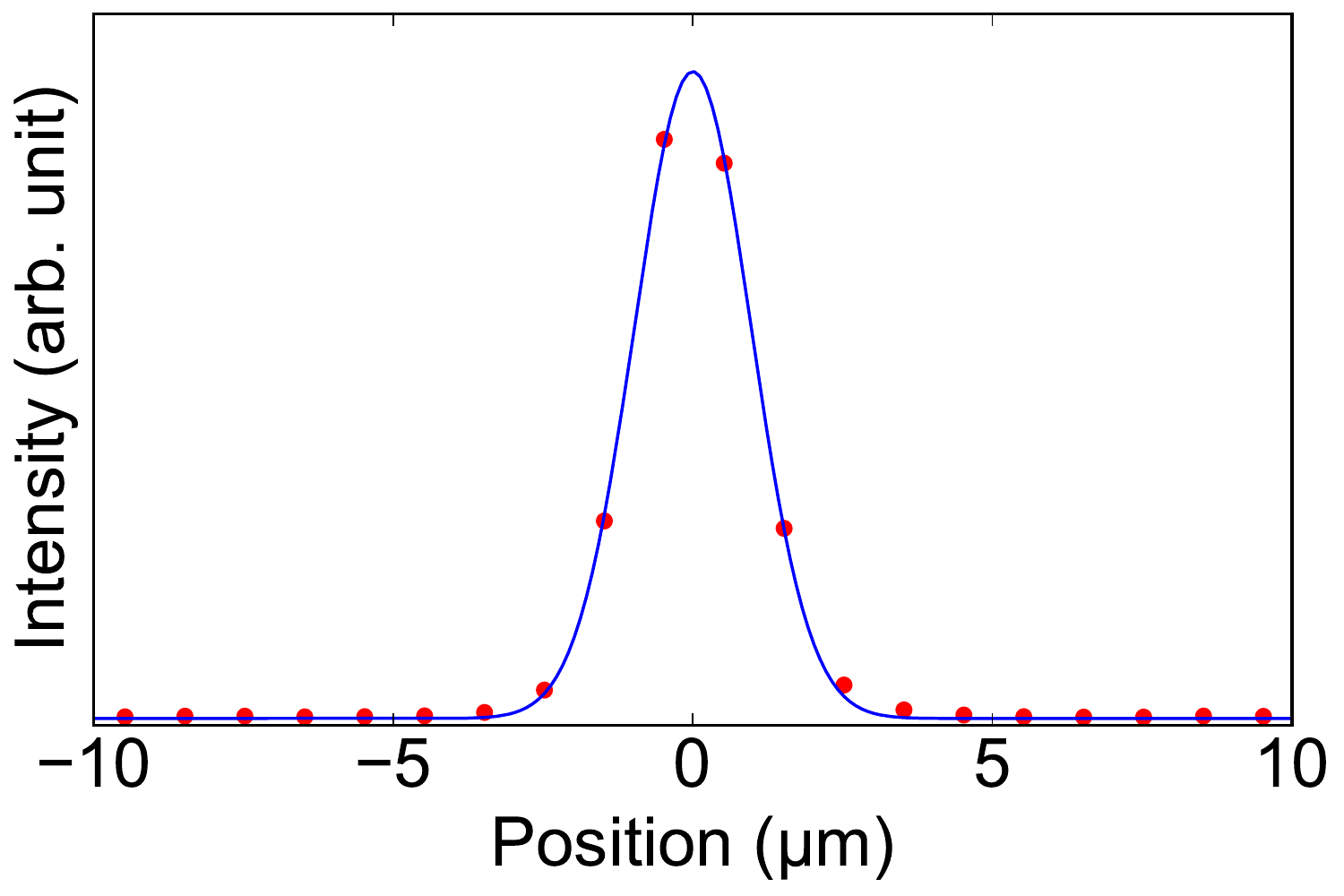}
\includegraphics[width=1.5in]{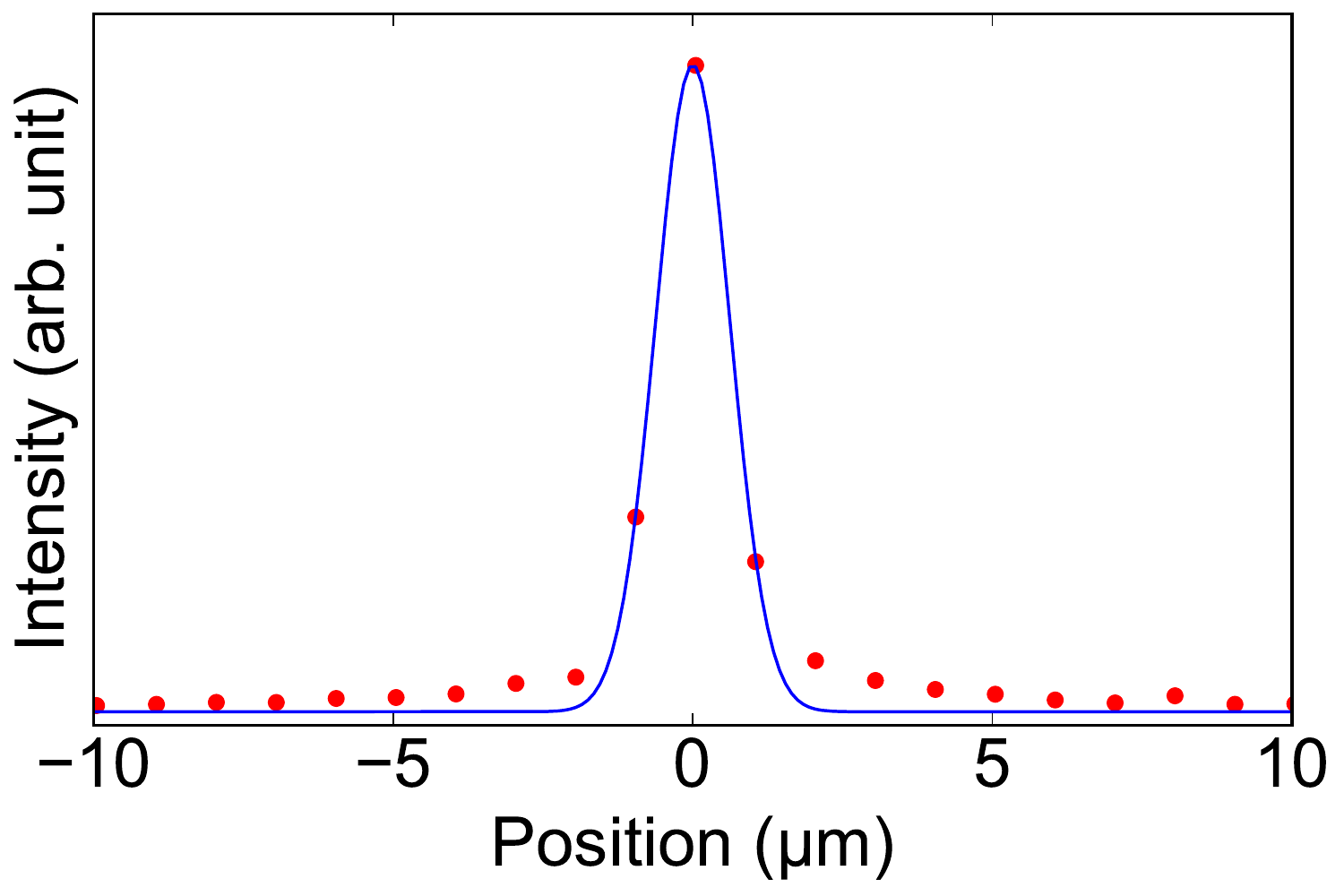}
\caption{Intensity profile on the focusing plane of a narrowband laser, recorded by a line camera. A large input beam size $w_i=12.1\;\text{mm}$ is used in the right panel, and causes the focal spot to be less compact. While in the left, a permissible beam size $w_i=6.8\;\text{mm}$ is used, and the focal spot profile is fairly close to Gaussian.}\label{fig_spot}
\end{figure}

In Fig.~\ref{fig_spot} we measure the profile of the focal spot for two different input beam sizes: $w_i$ = 6.8 mm (left) and 12.1 mm (right). (The input beam size is controlled by the fiber collimation lens.) In both cases, a compact Gaussian-like peak is observed. But the profile due to the larger beam also has additional tails next to the central peak. As the extra feature contains non-negligible intensity, it significantly degrades the resolution even when a larger beam is supposed to be more tightly focused. In spectral filtering, the tail results in a low extinction cutoff.

The origin of the tail is mostly from the transverse spherical aberration, which grows rapidly as the cube of the aperture size. Spherical aberration limits the clear aperture of an optics for tight focusing. From the above result, we learned that any input beam with $w_i<6.8\;\text{mm}$ should acquire only modest spherical aberration. Our homebuilt doublet lens is then satisfactory in this project as the broadband UV beam size (refer to Sec.~\ref{sec:light_source}) fits into its clear aperture.

\subsection{Other aberrations}\label{sec_evaluate2}
\begin{figure}
(a)\newline
\includegraphics[width=3in]{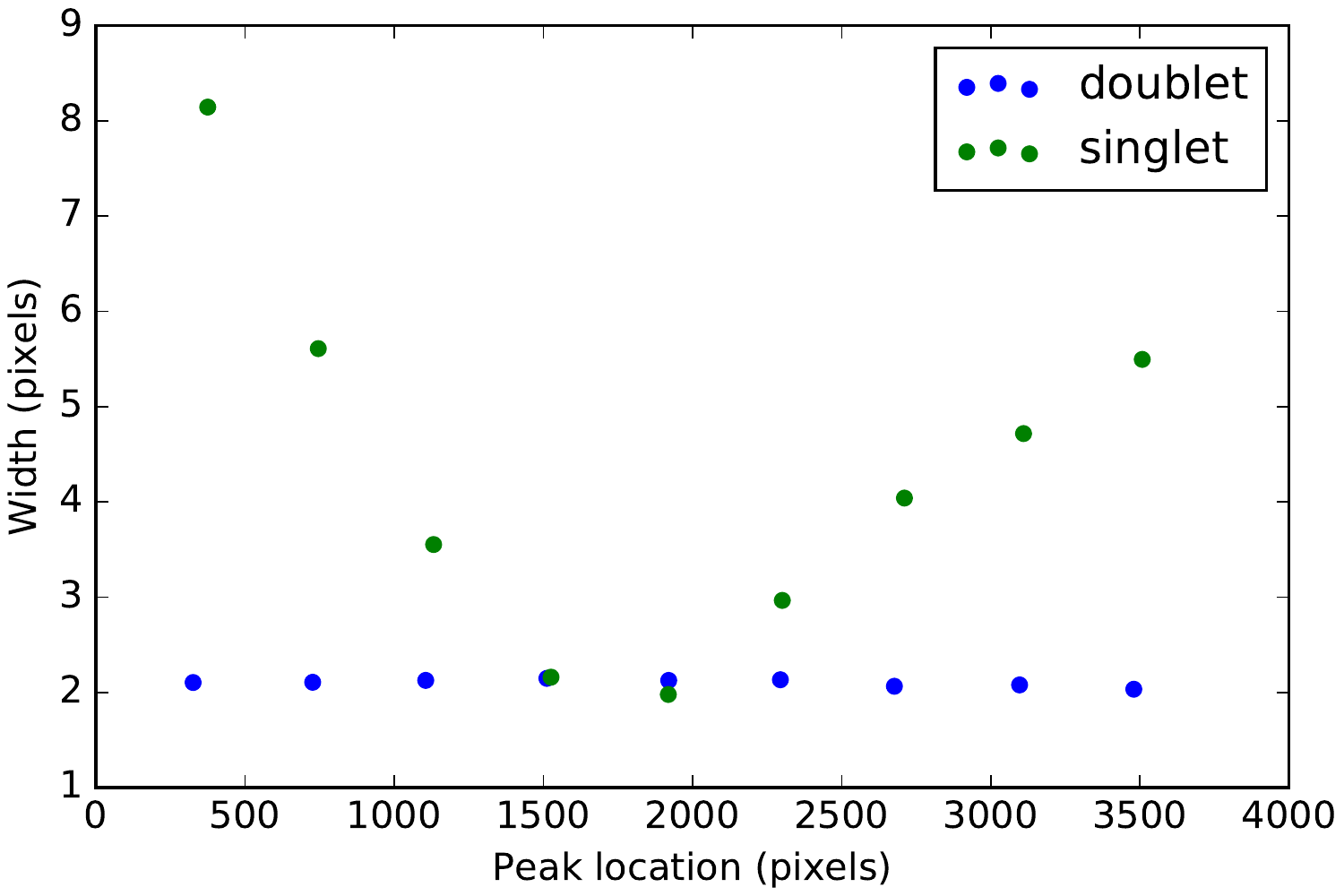}
\newline
(b)\newline
\includegraphics[width=3in]{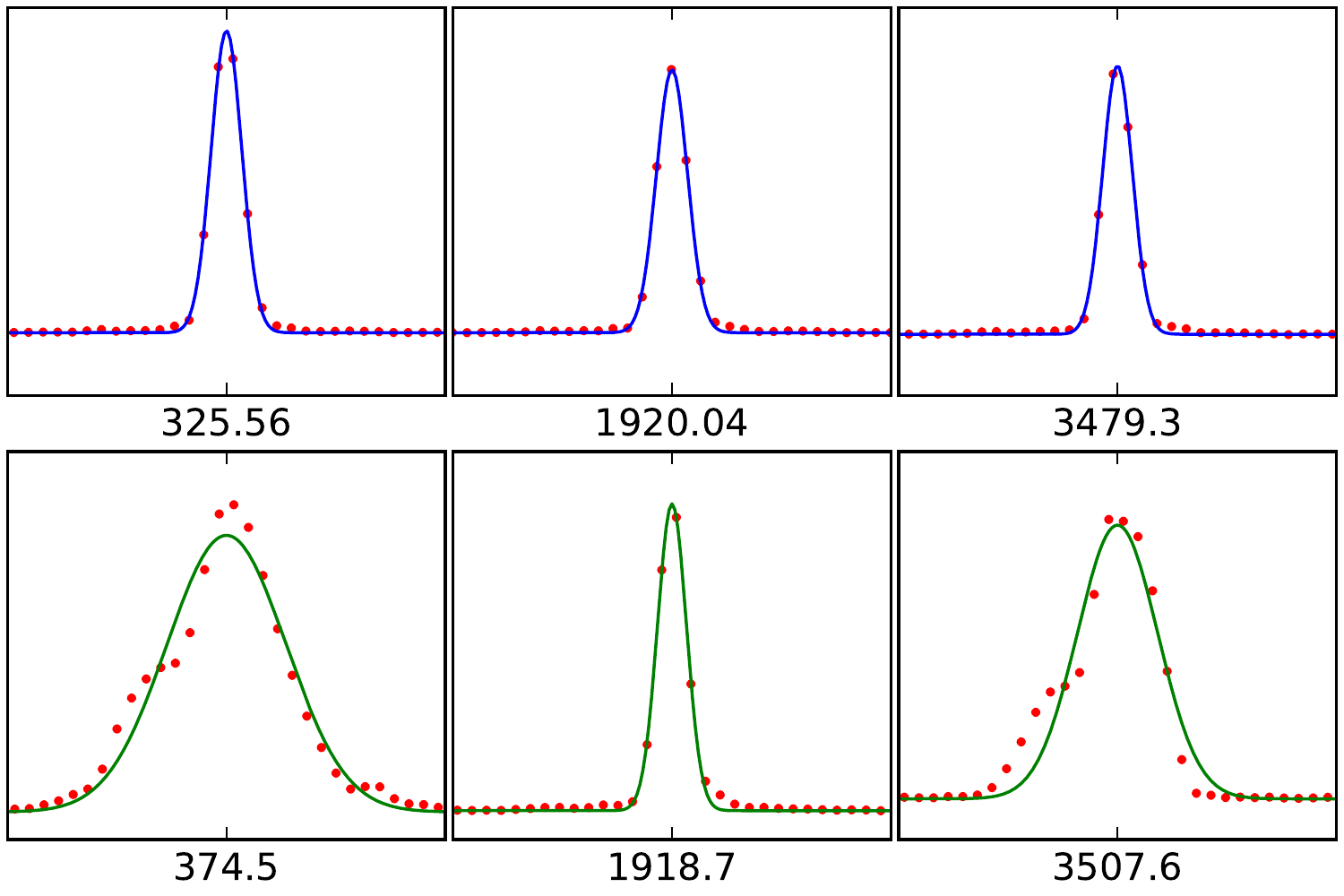}
\caption{(a) Consistency of spectrometer resolution over its measurement range. In this test, we simulate different narrowband laser wavelengths by rotating the grating, as if the diffraction angle is changed due to another wavelength. (b) Spot profiles for data \#1 (left-most), \#5 (middle), and \#9 (right-most) from the data sequence in (a). The horizontal axis is the pixel number on the linear camera. The spots in the upper row are focused by the doublet lens and those in the bottom row are focused by the singlet lens. The width of each plot is 30 pixels.}\label{fig_spec_calibration}
\end{figure}

In this section, we exam the focusing behavior for spectral components that access the lens off-center. A different narrowband spectral component can be obtained by changing the wavelength of the CW laser. It, however, requires a fancy laser system to modulate the laser wavelength over several nanometers. Therefore, to simulate tilted beams due to different wavelengths, we rotated the grating instead. We measured the focal spot profile for various diffracted beam pointing in Fig.~\ref{fig_spec_calibration}. For comparison, we performed the same measurement with a singlet lens as well. In (a) we have the peak width on the camera versus different tilt; in (b) some sample spot profiles at various tilts are presented. The focal spot size stays around 16 $\upmu$m over an approximately $\pm 2^\circ$ tilt with the doublet lens, while it varies quite a lot with the singlet lens. The result shows us that we can expect a relatively consistent resolution over the entire spectral range both in the spectrometer and the spectral filtering setup.

From the same data we also calibrate the spectrometer's linear dispersion. We first read the peak locations for each tilt from Fig.~\ref{fig_spec_calibration}(b) and fit to a straight line. Then we computed the angular dispersion due to the grating in this spectrometer. With both pieces of information, we obtained the calibration 7.2 GHz/pixel. In addition, the focal spot size converts to a 14 GHz spectral resolution in our spectrometer.

\subsection{spectral filtering}
\begin{figure}
\includegraphics[width=3in]{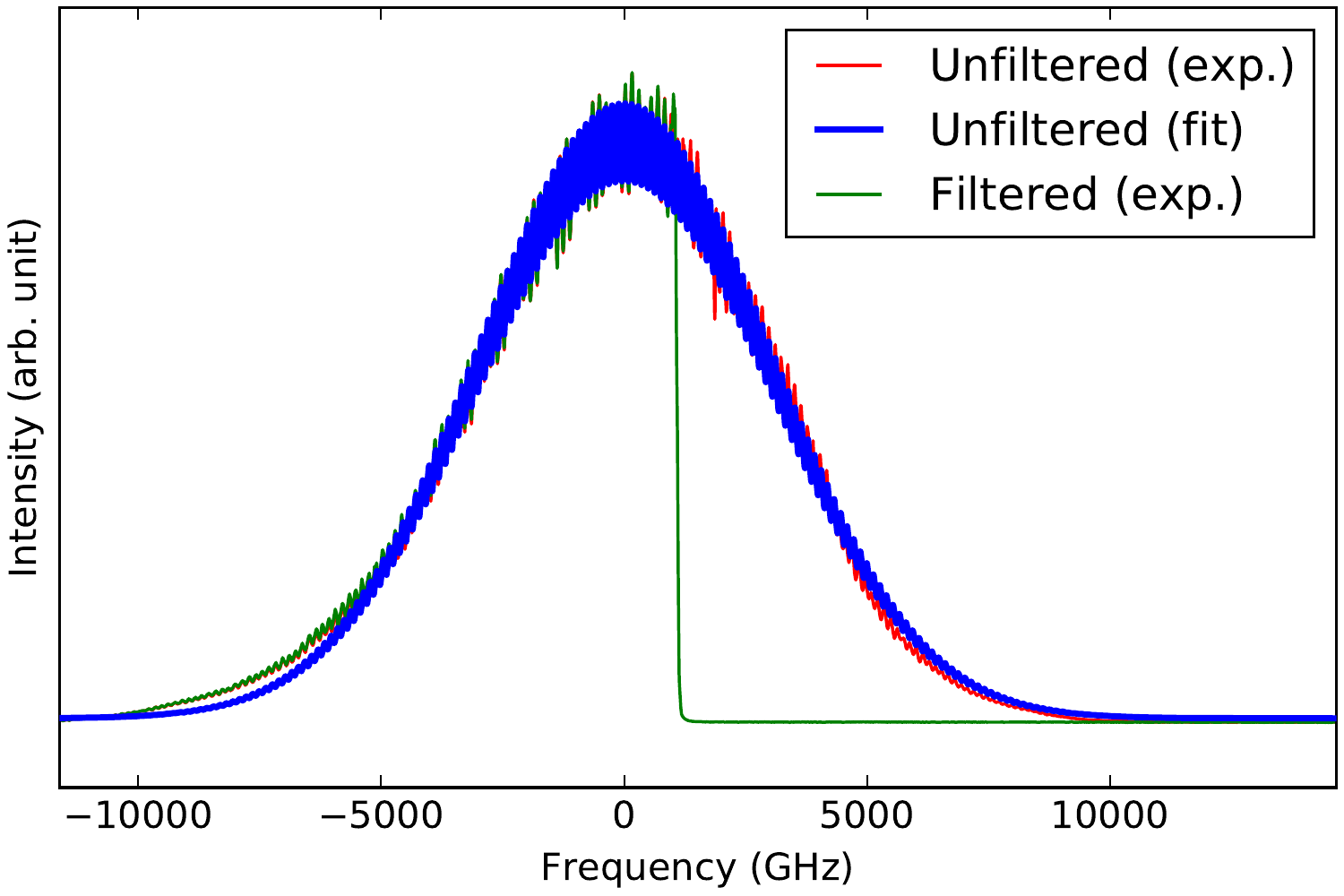}
\includegraphics[width=3in]{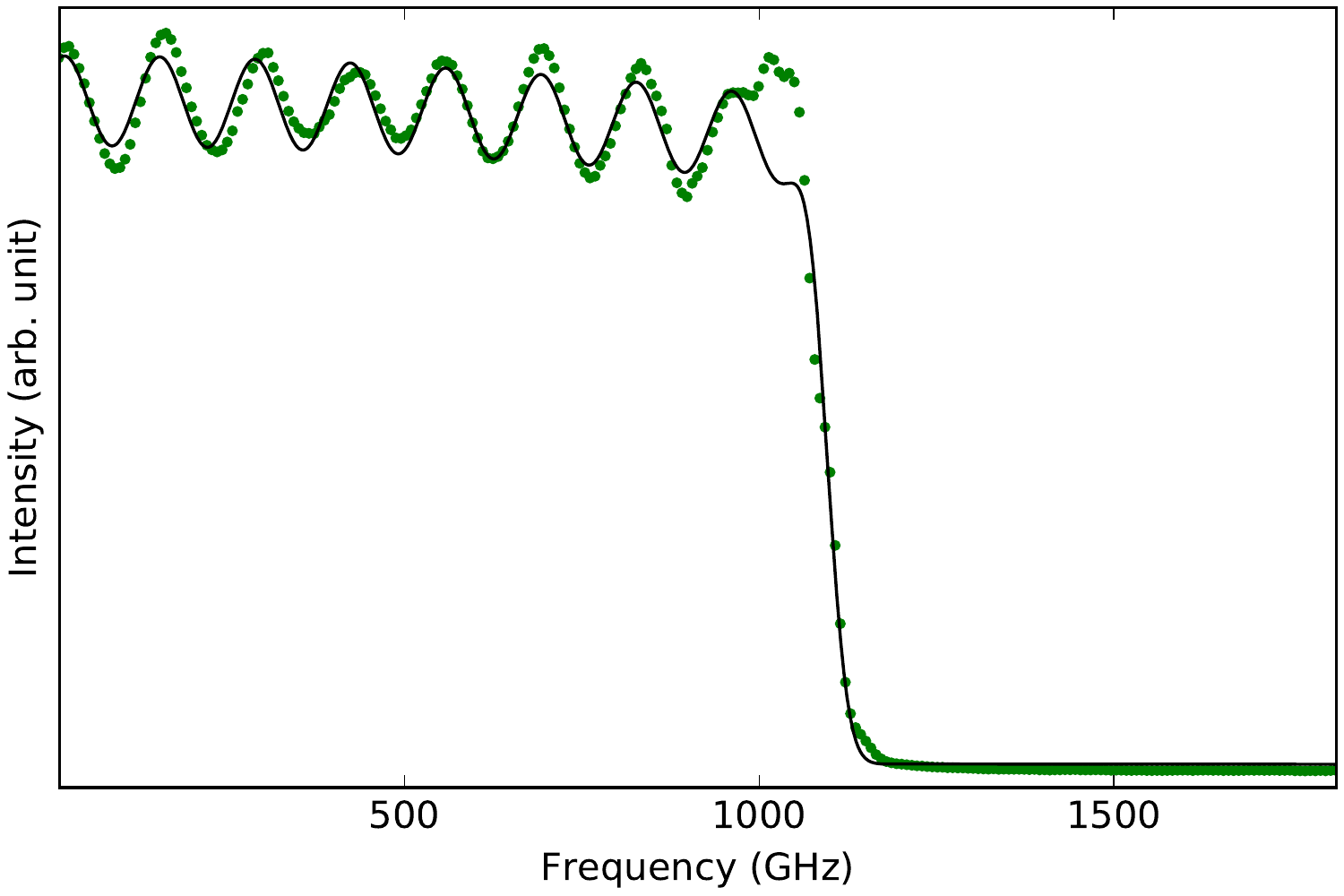}
\caption{Spectrum of the original and filtered light.}\label{fig_maitai_spec}
\end{figure}

With the spectrometer and the focusing lens characterized, we now turn the attention to the overall performance of our spectral filtering. We sent a few percent of the output light power into our spectrometer for diagnosis. In Fig.~\ref{fig_maitai_spec} we see the output spectrum with (green line) and without (red line) masking. The unfiltered spectrum has a Gaussian-like overall shape. The origin of the fringes on top of the profile is not yet clear to us. However, we were able to see the same fringes right after the UV generation crystal, therefore our best guess is it might come from the second harmonic generation process. Without further knowledge of the fringes, we simply model the feature as a sinusoidal modulation to the Gaussian function:
\begin{equation}
A\exp \left(-(\frac{\nu-\nu_c}{\Delta\nu})^2\right)\times \left[1+h\sin\frac{2\pi(\nu-\nu_p)}{\Omega}\right]
\end{equation}
where $\nu_c$ is the broadband laser center frequency, $\Delta\nu$ is the laser bandwidth, $\Omega$ is the fringe period, $\nu_p$ is used to adjust the fringe's offset, and $h$ is the modulation index. The unfiltered spectrum has $\Delta\nu = 7.1\;\text{THz}$ full-width-half-maximum bandwidth, and the period of the fringes is around $\Omega=150\;\text{GHz}$.

For the cutoff, we model it with an error function
\begin{equation}
g(\nu, \nu_0) = g(\nu-\nu_0) = \frac{1}{2}[\text{erf} (\frac{\nu-\nu_0}{\sqrt{2} \delta\nu})+1]
\end{equation}
where $\nu_0$ is the cutoff frequency and $\delta\nu$ is the cutoff width. Positive $\delta\nu$ represents a high-pass mask and negative $\delta\nu$ is low-pass. The error function is a natural choice in our analysis as the edge of the mask partially blocks a Gaussian beam. Fig.~\ref{fig_maitai_spec}(b) shows a typical filtered spectrum near the cutoff with our model applied to the data. In this demonstration, the cutoff width $\delta\nu$ is 48 GHz. Compared to the expected diffraction-limited resolution, this result is 33\% worse.

\begin{figure}
\includegraphics[width=3in]{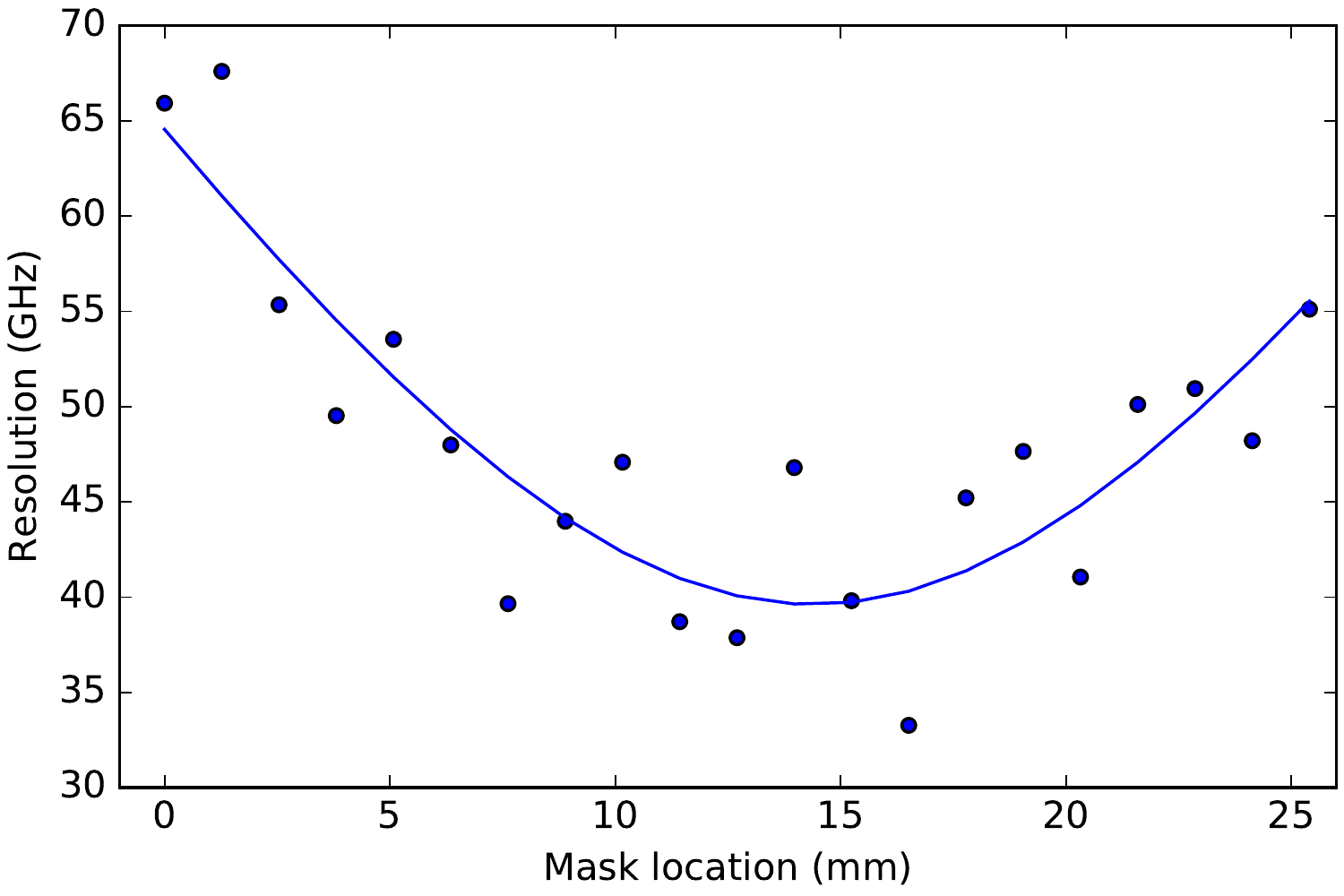}
\caption{Cutoff resolution versus mask longitudinal position.}\label{fig_width_long}
\end{figure}

In Fig.~\ref{fig_width_long} we measure the width $\delta\nu$ for different mask longitudinal locations $z$. The result is modeled by
\begin{equation}
\delta\nu(z) = \delta\nu_0 \sqrt{1+(\frac{z-z_0}{z_R})^2}
\end{equation}
which resembles the equation for Gaussian beam width evolution. The range $z_R=\pm 11.2\;\text{mm}$ is considered the alignement tolerance of the mask.

\begin{figure}
\includegraphics[width=3in]{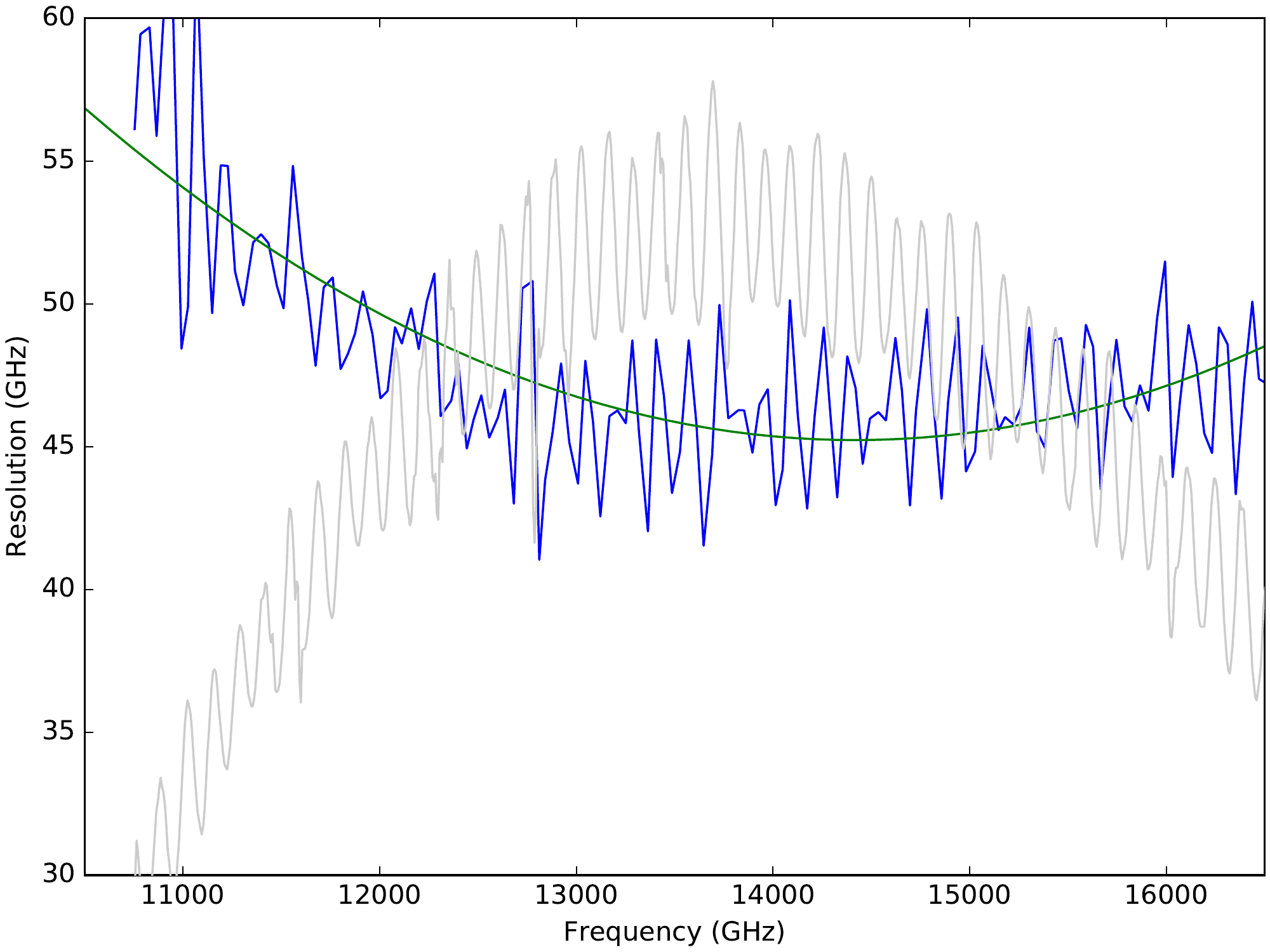}
\caption{cutoff resolution versus different cutoff frequency. }\label{fig_cutoff}
\end{figure}

We also measured the cutoff behavior at different cutoff frequencies to test the consistency over the spectrum. We found the cutoff frequency is related to the blade's transverse position by $1.6\;\text{GHz}/\upmu\text{m}$; this is also a measurement of the linear dispersion of our spectral filtering setup. In Fig.~\ref{fig_cutoff} we plot the cutoff width as a function of cutoff frequency. We notice that the cutoff width varies rapidly across the laser spectrum, and the variation coincides with the spectral fringes from the unfiltered light, shown in the gray line. As we currently have no plausible explanation for the fringes, we cannot understand the correlation either. Regardless, by eye-averaging the result, we find the cutoff resolution is worsened by around 20\% when away from the central wavelength, which can be attributed to the optical aberration.

\section{Improvement}
We have demonstrated that the focusing lens is the key component in both the spectrometer and the spectral filtering setup. In this project, we used standard optics components available from common vendors, which significantly limited the ability of aberration correction. The immediate also less expensive solution is to build a doublet lens with proper choices of lens materials and surface curvatures. By doing optical ray tracing simulations, we, however, found that, for doublet lenses utilizing spherical surfaces, the largest clear aperture diameter is somehow around 20 mm. The performance seems to be limited by the number of tuning variables in lens design. That is, if one wants to work with larger input beam size, it is necessary to use aspheric optics.

The input beam shape is another factor that might need more studying. While SHG of a femtosecond laser is quite simple, we have observed evidence of beam shape degradation. The obvious one is astigmatism where the UV beam presents an elliptical shape. In addition, the UV beam is accompanied by a halo; it is not clear to us how this scattered light behaves in the pulse shaping setup and affects the filtering resolution.

To ultimately reach close to the diffraction limit, wavefront distortion caused by, e.g., optics surface irregularity, is yet another obstacle to be studied and overcome.

Finally, regarding the mask, a razor blade serves as an adequate bandpass filter for our molecular cooling experiment. For comprehensive modulation, a spatial light modulator (SLM) is a better option. SLMs with good resolution and dynamic range have become available in the UV band. Options to be considered for our future experiment include liquid crystal\citep{Tanigawa:2009aa}, acousto-optic modulator\citep{Roth:2005aa,Pearson:2007aa}, and micro-mirror array\citep{Rondi:2009aa}.

\section{Summary}
We studied using ultrashort pulse shaping techniques to perform spectral filtering for molecular rotational cooling. We investigated how each component in the implementation affects the filtering resolution and extinction ratio, which are particularly important in the molecular cooling experiment. With standard stocked UV optics, we aligned the setup and demonstrated better than 100 GHz filtering resolution with close to error function cutoff behavior. Our result is currently limited by the clear aperture of our optics but is by far the best in UV with a 4-f configuration to our knowledge. Moreover, our spectral filtering will provide sufficient control of the broadband laser spectrum for our ongoing SiO$^+$ cooling project.

\section*{Acknowledgement}
We thank John Marko for assisting us examining the razor blades on his microscope as shown in Fig.~\ref{fig_blade_image}. We are also grateful for Marcos Dantus and Daniel Leaird's comments and advice on this work.


\begin{thebibliography}{10}

\bibitem{sardesai1998femtosecond}
HP~Sardesai, C-C Chang, and AM~Weiner.
\newblock A femtosecond code-division multiple-access communication system test
  bed.
\newblock {\em Journal of Lightwave Technology}, 16(11):1953--1964, 1998.

\bibitem{Pastirk:2006aa}
Igor Pastirk, Bojan Resan, Alan Fry, John MacKay, and M~Dantus.
\newblock No loss spectral phase correction and arbitrary phase shaping of
  regeneratively amplified femtosecond pulses using miips.
\newblock {\em Optics Express}, 14(20):9537--9543, 2006.

\bibitem{goswami2003optical}
Debabrata Goswami.
\newblock Optical pulse shaping approaches to coherent control.
\newblock {\em Physics Reports}, 374(6):385--481, 2003.

\bibitem{Sauer:2007aa}
Franziska Sauer, Andrea Merli, Ludger W{\"o}ste, and Albrecht Lindinger.
\newblock High resolution coherent control measurements on krb.
\newblock {\em Chemical physics}, 334(1):138--143, 2007.

\bibitem{Lien:2014aa}
Chien-Yu Lien, Christopher~M Seck, Yen-Wei Lin, Jason H.~V. Nguyen, David~A.
  Tabor, and Brian~C. Odom.
\newblock Broadband optical cooling of molecular rotors from room temperature
  to the ground state.
\newblock {\em Nat Commun}, 5, 09 2014.

\bibitem{Lien:2011aa}
Chien-Yu Lien, Scott~R. Williams, and Brian Odom.
\newblock Optical pulse-shaping for internal cooling of molecules.
\newblock {\em Physical Chemistry Chemical Physics}, 13(42):18825--18829, 2011.

\bibitem{Thurston:1986aa}
R.~Thurston, J.~Heritage, A.~Weiner, and W.~Tomlinson.
\newblock Analysis of picosecond pulse shape synthesis by spectral masking in a
  grating pulse compressor.
\newblock {\em IEEE Journal of Quantum Electronics}, 22(5):682--696, May 1986.

\bibitem{Monmayrant:2004aa}
Antoine Monmayrant and B{\'e}atrice Chatel.
\newblock New phase and amplitude high resolution pulse shaper.
\newblock {\em Review of Scientific Instruments}, 75:2668--2671, 2004.

\bibitem{Stobrawa:2001aa}
G~Stobrawa, M~Hacker, Th~Feurer, D~Zeidler, M~Motzkus, and F~Reichel.
\newblock A new high-resolution femtosecond pulse shaper.
\newblock {\em Applied Physics B}, 72(5):627--630, 2001.

\bibitem{Lee:2006aa}
Ghang-Ho Lee, Shijun Xiao, and Andrew~M Weiner.
\newblock Optical dispersion compensator with> 4000-ps/nm tuning range using
  (vipa) and spatial light modulator (slm).
\newblock {\em IEEE photonics technology letters}, 18(17):1819, 2006.

\bibitem{Supradeepa:08}
V.R. Supradeepa, Chen-Bin Huang, Daniel~E. Leaird, and Andrew~M. Weiner.
\newblock Femtosecond pulse shaping in two dimensions: Towards higher
  complexity optical waveforms.
\newblock {\em Opt. Express}, 16(16):11878--11887, Aug 2008.

\bibitem{Weber:2010aa}
S{\'e}bastien Weber, M~Barthelemy, and B~Chatel.
\newblock Direct shaping of tunable uv ultra-short pulses.
\newblock {\em Applied Physics B}, 98(2-3):323--326, 2010.

\bibitem{Monmayrant:2010aa}
Antoine Monmayrant, S{\'e}bastien Weber, and B{\'e}atrice Chatel.
\newblock A newcomer's guide to ultrashort pulse shaping and characterization.
\newblock {\em Journal of Physics B: Atomic, Molecular and Optical Physics},
  43(10):103001, 2010.

\bibitem{Weiner:2011aa}
Andrew~M. Weiner.
\newblock Ultrafast optical pulse shaping: A tutorial review.
\newblock {\em Optics Communications}, 284(15):3669 -- 3692, 2011.
\newblock Special Issue on Optical Pulse Shaping, Arbitrary Waveform
  Generation, and Pulse Characterization.

\bibitem{Paye:1995aa}
J{\'e}r{\^o}ome Paye and Arnold Migus.
\newblock Space--time wigner functions and their application to the analysis of
  a pulse shaper.
\newblock {\em JOSA B}, 12(8):1480--1490, 1995.

\bibitem{Wefers:1996aa}
Marc~M Wefers and Keith~A Nelson.
\newblock Space-time profiles of shaped ultrafast optical waveforms.
\newblock {\em IEEE Journal of Quantum Electronics}, 32(1):161--172, 1996.

\bibitem{Zhang:1998aa}
Jing-yuan Zhang, Jung~Y Huang, H~Wang, KS~Wong, and GK~Wong.
\newblock Second-harmonic generation from regeneratively amplified femtosecond
  laser pulses in {BBO} and {LBO} crystals.
\newblock {\em JOSA B}, 15(1):200--209, 1998.

\bibitem{Ghotbi:2004aa}
Masood Ghotbi and M~Ebrahim-Zadeh.
\newblock Optical second harmonic generation properties of {$BiB_3O6$}.
\newblock {\em Optics Express}, 12(24):6002--6019, 2004.

\bibitem{Ghotbi:2004ab}
M~Ghotbi, M~Ebrahim-Zadeh, A~Majchrowski, E~Michalski, and IV~Kityk.
\newblock High-average-power femtosecond pulse generation in the blue using
  $bib_3o_6$.
\newblock {\em Optics letters}, 29(21):2530--2532, 2004.

\bibitem{Tanigawa:2009aa}
Takashi Tanigawa, Yu~Sakakibara, Shaobo Fang, Taro Sekikawa, and Mikio
  Yamashita.
\newblock Spatial light modulator of 648 pixels with liquid crystal transparent
  from ultraviolet to near-infrared and its chirp compensation application.
\newblock {\em Optics letters}, 34(11):1696--1698, 2009.

\bibitem{Roth:2005aa}
M~Roth, M~Mehendale, A~Bartelt, and H~Rabitz.
\newblock Acousto-optical shaping of ultraviolet femtosecond pulses.
\newblock {\em Applied Physics B}, 80(4-5):441--444, 2005.

\bibitem{Pearson:2007aa}
Brett~J Pearson and Thomas~C Weinacht.
\newblock Shaped ultrafast laser pulses in the deep ultraviolet.
\newblock {\em optics express}, 15(7):4385--4388, 2007.

\bibitem{Rondi:2009aa}
Ariana Rondi, J{\'e}r{\^o}me Extermann, Luigi Bonacina, SM~Weber, and J-P Wolf.
\newblock Characterization of a {MEMS}-based pulse-shaping device in the deep
  ultraviolet.
\newblock {\em Applied Physics B}, 96(4):757--761, 2009.

\end{thebibliography}

\end{document}